\documentclass{article}

\usepackage{microtype}
\usepackage{graphicx}
\usepackage{booktabs} 

\usepackage{hyperref}

\usepackage[accepted]{include/icml2019/icml2019}

\usepackage{amsfonts}
\usepackage{amsmath}
\usepackage{amsthm}
\usepackage{dsfont}
\usepackage{mathtools}
\usepackage{multirow}
\usepackage{scalerel}
\usepackage{subcaption}  

\newcommand\Cite[1]{ \cite{#1}}
\newcommand\OnlineCite[1]{\citeauthor{#1} (\citeyear{#1})}
\newcommand\pubB{\mathcal{B}}

\DeclareMathAlphabet{\mathsfsl}{OT1}{cmss}{m}{sl}

\renewcommand{\phi}{\varphi}

\newcommand{\pidet}{\overset{\lower.1em\hbox{\scaleto{\Delta}{2.5pt}}}{\pi}}
\newcommand{\finalsteps}{16.3}
\newcommand{\finalperformance}{24.174}
\newcommand{\finalperfectgames}{58.6}

\newcommand{\fei}{f[i]}
\newcommand{\fej}{f[j]}

\newcommand{\gei}{g[i]}
\newcommand{\gej}{g[j]}

\newcommand{\ftei}{f_t[i]}
\newcommand{\fmi}{f[\scalebox{0.5}[1.0]{\( - \)}i]}
\newcommand{\gmi}{g[\scalebox{0.5}[1.0]{\( - \)}i]}

\newcommand{\Expect}{\operatorname{\mathbb{E}}}

\newcommand{\myvec}[1]{\mathbf{#1}}

\newcommand{\vu}{\myvec{u}}

\newcommand{\be}{\begin{equation}}
\newcommand{\ee}{\end{equation}}
\newcommand{\bea}{\begin{eqnarray}}
\newcommand{\eea}{\end{eqnarray}}
\newcommand{\beaa}{\begin{eqnarray*}}
\newcommand{\eeaa}{\end{eqnarray*}}

\DeclareMathAlphabet{\mathpzc}{OT1}{pzc}{m}{n}

\newcommand{\mathbbm}[1]{\mathds{#1}}

\icmltitlerunning{Bayesian Action Decoder}

\begin{document}

\twocolumn[
\icmltitle{Bayesian Action Decoder for Deep Multi-Agent Reinforcement Learning}



\icmlsetsymbol{equal}{*}

\begin{icmlauthorlist}
\icmlauthor{Jakob N. Foerster}{equal,ox,workdone}
\icmlauthor{H. Francis Song}{equal,dm}
\icmlauthor{Edward Hughes}{dm}
\icmlauthor{Neil Burch}{dm}
\icmlauthor{Iain Dunning}{dm}
\icmlauthor{Shimon Whiteson}{ox}
\icmlauthor{Matthew M. Botvinick}{dm}
\icmlauthor{Michael Bowling}{dm}
\end{icmlauthorlist}

\icmlaffiliation{ox}{University of Oxford, UK}
\icmlaffiliation{dm}{DeepMind, London, UK}
\icmlaffiliation{workdone}{Work done at DeepMind. JF has since moved to Facebook AI Research, Menlo Park, USA.}

\icmlcorrespondingauthor{Jakob Foerster}{jnf@fb.com}
\icmlcorrespondingauthor{Francis Song}{songf@google.com}

\icmlkeywords{Machine Learning, ICML}

\vskip 0.3in
]



\printAffiliationsAndNotice{\icmlEqualContribution} 


\begin{abstract}
When observing the actions of others, humans make inferences about why they acted as they did, and what this implies about the world; humans also use the fact that their actions will be interpreted in this manner, allowing them to act informatively and thereby communicate efficiently with others. Although learning algorithms have recently achieved superhuman performance in a number of two-player, zero-sum games, scalable multi-agent reinforcement learning algorithms that can discover effective strategies and conventions in complex, partially observable settings have proven elusive. We present the \emph{Bayesian action decoder} (BAD), a new multi-agent learning method that uses an approximate Bayesian update to obtain a public belief that conditions on the actions taken by all agents in the environment. BAD introduces a new Markov decision process, the \emph{public belief MDP}, in which the action space consists of all deterministic partial policies, and exploits the fact that an agent acting only on this public belief state can still learn to use its private information if the action space is augmented to be over all partial policies mapping private information into environment actions. The Bayesian update is closely related to the \emph{theory of mind} reasoning that humans carry out when observing others' actions. We first validate BAD on a proof-of-principle two-step matrix game, where it outperforms policy gradient methods; we then evaluate BAD on the challenging, cooperative partial-information card game Hanabi, where, in the two-player setting, it surpasses all previously published learning and hand-coded approaches, establishing a new state of the art.

\end{abstract}


\section{Introduction}
In multi-agent reinforcement learning (RL), agents must learn to act in an environment that contains multiple learning agents, often under partial observability \cite{Littman1994}. In recent years, a variety of deep RL methods have been adapted to this setting \cite{foerster2016learning,lowe2017multi,Perolat2017,Jaderberg2018}. In the particular case of cooperative, partially observable multi-agent settings, a key challenge is to discover communication protocols while simultaneously learning policies. Such protocols are essential for many real-world tasks where agents must interact and communicate seamlessly with other agents. %

State-of-the-art deep RL methods for learning communication protocols mostly use backpropagation across a communication channel \cite{Sukhbaatar2016,foerster2016learning}.  This approach has two limitations. First, it can only be applied to \emph{cheap-talk} channels in which the communication action has no effect on the environment. 
Second, it misses the conceptual connection between communication and reasoning over the beliefs of others, which is known to be important to how humans learn to communicate \cite{Grice1975,Frank2012}. %

\begin{figure*}[t]
 	\centering
		\includegraphics[width=0.9\linewidth]{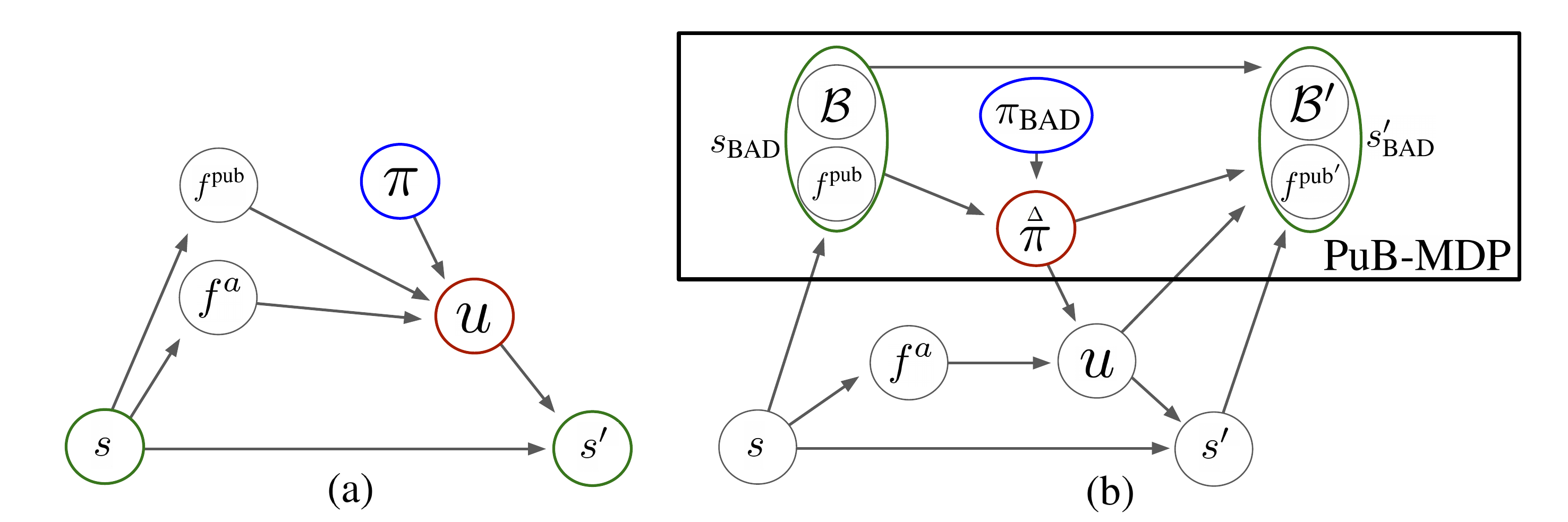}
 	\caption{a) In an MDP the action $u$ is sampled from a policy $\pi$ that conditions on the state features (here separated into $f^\text{pub}$ and $f^a$). The next state is sampled from $P(s' | s, u)$. 
 	 	b) In a PuB-MDP, public features $f^\text{pub}$ generated by the environment and the public belief together constitute the Markov state $s_\text{BAD}$. The `action' sampled by the BAD agent is in fact a deterministic partial policy $\pidet \sim \pi_{\text{BAD}}(\pidet |s_\text{BAD})$ that maps from private observations $f^a$ to actions.
Only the acting agent  observes $f^a$ and deterministically computes $u = \pidet(f^a)$. $u$ is provided to the environment, which transitions to state $s'$ and produces the new observation $f^\text{pub$'$}$. BAD then uses the public belief update to compute a new belief $\pubB'$ conditioned on $u$ and $\pidet$ (Equation~\ref{eq:belief_update}), thereby completing the state transition.}
 	\label{fig:fig_1}
 \end{figure*}

A well-known domain that highlights these challenges is Hanabi, a popular, fully cooperative card game of incomplete information that is difficult even for humans (Hanabi won the \emph{Spiel des Jahres} award in 2013).  A distinguishing feature of the game is that players see everyone's hands but their own and must find effective conventions for communication to succeed.  Since there is no cheap-talk channel, most recent methods for emergent communication are inapplicable, necessitating a novel approach. Because of these unique features, Hanabi has recently been proposed as a new benchmark for multi-agent learning~\cite{bard2019hanabi}.

The goal in Hanabi is to play a legal sequence of cards and, to aid this process, players are allowed to give each other hints indicating which cards are of a specific rank or colour.  These hints have two levels of semantics.  The first level is the surface-level content of the hint, which is grounded in the properties of the cards that they describe.  This level of semantics is independent of any possible intent of the agent in providing the hint, and would be equally meaningful if provided by a random agent. For example, knowing which cards are of a specific colour often does not indicate whether they can be safely played or discarded. 

A second level of semantics arises from information contained in the actions themselves, i.e., the very fact that an agent decided to take a particular action and not another, rather than the information resulting from the state transition induced by the action. This is essential to the formation of conventions and to discovering good strategies in Hanabi. 

To address these challenges, we propose the \emph{Bayesian action decoder} (BAD), a novel multi-agent RL algorithm for discovering effective communication protocols and policies in cooperative, partially observable multi-agent settings. Inspired by the work of~\citet{Nayyar2013}, BAD uses all publicly observable features in the environment to compute a  \emph{public belief} over the players' private features. %
This effectively defines a new Markov process, the \emph{public belief Markov decision process} (PuB-MDP), in which the action space is the set of deterministic partial policies, parameterised by deep neural networks, that can be sampled for a given public state.
 By acting in the space of deterministic partial policies that map from private observations into environment actions, an agent acting only on this public belief state can still learn an optimal policy. 
 Using approximate, factorised Bayesian updates and deep neural networks, we show for the first time how a method using the public belief of ~\citet{Nayyar2013}, can scale to large state spaces and allow agents to carry out a form of counterfactual reasoning.

When an agent observes the action of another agent, the public belief is updated by sampling a set of possible private states from the public belief and filtering for those states in which the teammate chose the observed action. This process is closely related to the kind of \emph{theory of mind} reasoning that humans routinely undertake \cite{Baker2017}. Such reasoning seeks to understand why a person took a specific action among several, and what information this contains about the distribution over private observations.

We experimentally validate an exact version of BAD on a simple two-step matrix game, showing that it outperforms policy gradient methods. We then apply an approximate version to Hanabi, where BAD achieves
an average score of \finalperformance{} points in the two-player setting, surpassing the best previously published results for learning agents by around 9 points and approaching the best known performance of 24.9 points for (cheating) open-hand gameplay. BAD thus establishes a current state-of-the-art on the Hanabi Learning Environment~\cite{bard2019hanabi} for the two player self-play setting. We further show that the beliefs obtained via Bayesian reasoning have 40\% less uncertainty over possible hands than those using only grounded information.

\section{Background and Setting}
\label{sec:background}
Consider a partially observable multi-agent environment with $A$ agents. At time $t$ each agent $a$ takes action $u^a_t$ sampled from policy $\pi^a( u^a |  \tau^a_t)$, where $\tau^a_t$ is its action-observation history  $\tau^a_t = \{o^a_0, u^a_0, .. ,o^a_t\}$. Here $o^a_t$ are the observations of agent $a$ at time $t$, which is given by the observation function $O(a, s_t)$ in state $s_t$. The next Markov state $s_{t+1}$ of the environment is produced by the transition function $P(s_{t+1} |  s_t, \vu_t)$, which conditions on the joint action $\vu_t = \{u^1_t, .., u^A_t\}$, where $u^a_t \in \mathcal{U}$. In the fully cooperative setting considered here, each agent receives a per-timestep team reward $r_{t+1}(s_t, \vu_t)$ that depends on the last state and last joint action. We allow centralised training but require decentralised execution, from which follows that the policies $\pi^a$ are known to all agents. This setting can be formalised as a Dec-POMDP~\cite{oliehoek2012decentralized}.

The goal of multi-agent RL is to find a set of agent policies $\{ \pi^a \}_{a=1,\ldots,A}$ that maximise the total expected return per episode $J =  \Expect_{\tau \sim P(\tau | \pi^a )} \big[{\sum_t  \gamma^t r_t} \big]$, where $\gamma$ is the discount factor. 
In deep RL, optimisation involves training neural networks that represent policies and value functions.
In partially observable settings, the networks are typically recurrent, e.g., LSTMs \cite{Wierstra2009}, as they can learn to represent a sufficient statistic of the action-observation history $\tau^a_t $ in the hidden activations.

Here we consider a setting where the Markov state $s_t$ consists of a set of discrete features $f_t$, composed of public features $f^\text{pub}_t$ and private features $f^\text{pri}_t$. The public features are common knowledge to all agents,
while private features are observable by at least one, but not all, of the agents. $f_t^{a}$ are the private features observable by agent $a$. We use the notation $f^\text{pri}_t[i]$ to indicate the $i$-th private state feature.
For example, in a typical card game the cards being played openly on the table are part of $f^\text{pub}_t$,  the cards held by each player are in  $f^\text{pri}_t $, $f^{a}_t$ contains the cards held by agent $a$, and $f^\text{pri}_t[i]$ corresponds to a specific card held by a specific player.
We assume that this separation of state features is common knowledge to all agents. 
An example of this separation for the case of an MDP is illustrated in Figure~\ref{fig:fig_1}a. Furthermore, while all our formalisms and methods can be extended to synchronous action settings, for simplicity we assume a turn-based setting with one agent acting per step.

\section{Method}
\label{sec:method}
Below we introduce the \emph{Bayesian Action Decoder} (BAD).  BAD scales the public belief of \citet{Nayyar2013} to large state spaces using factorised beliefs, an approximate Bayesian update, and sampled deterministic policies parameterised by deep neural networks.

\subsection{Public belief}
 In single-agent partially observable settings, it is clearly useful for an agent to maintain beliefs about the hidden environment state, since this is a sufficient statistic for its action-observation history \cite{Kaelbling1998}.  In multi-agent settings, however, it is not obvious what the beliefs should be over. It is not enough to maintain beliefs over the environment state alone, as other agents also have unobservable internal states.  In interactive POMDPs (I-POMDPs; \citeauthor{Gmytrasiewicz2005} \citeyear{Gmytrasiewicz2005}), agents model each other's beliefs, 
 beliefs over these beliefs, and so on, but this is often computationally intractable.
 
Fortunately, in our setting the common knowledge described above makes it possible to compute a \emph{public belief},~\cite{Nayyar2013}, that makes the recursion of I-POMDPs unnecessary. In our case the public belief $\pubB_t$ is the posterior over all of the private state features given only the public features, i.e.,
$\pubB_t = P( f^\text{pri}_t |f^\text{pub}_{\leq t})$, where $\leq t$ indicates history: $f^\text{pub}_{\leq t} = (f_0^\text{pub}, .. , f_t^\text{pub})$.  Because $\pubB_t$ conditions only on publicly available information, it can be computed independently by every agent via a common algorithm, yielding the same result for all agents.  Furthermore, since all agents know $f^\text{pub}$, it suffices for 
$\pubB_t$ to be a posterior over $f^\text{pri}$, not $f_t = \{f^\text{pri},f^\text{pub}\}$.

While the public belief avoids recursive reasoning, it is not obvious how it can be used to guide behaviour: agents that condition their actions only on the public belief will never exploit their private observations. As \citet{Nayyar2013} propose, we can construct a special \emph{public agent} whose policy $\pi_\text{BAD}$ conditions on the public observation and the public belief but which nonetheless can generate optimal behaviour.\footnote{$\pi_\text{BAD}$ conditions on the public observation because the public belief is a sufficient statistic for the public observation, but only over the private features. }
 This is possible because an action selected by $ \pi_\text{BAD}$ specifies a \emph{partial policy}, $\pidet : \{f^a\} \rightarrow \mathcal{U}$, for the acting agent, 
 deterministically mapping private observations to environment actions. The sampling of a deterministic partial policy also addresses a fundamental tension in using policy gradients to learn communication protocols, namely, differentiation and exploration require high-entropy policies, while communication requires low-entropy policies. By sampling in the space of deterministic policies, both can be achieved.

Intuitively, the public agent can be viewed as a third party that can observe only the public observation and belief.  While $\pi_\text{BAD}$ cannot observe the private state, it can tell each agent what to do for any private observation it might receive. Thus at each timestep, the public agent selects $\pidet $ based on $\pubB_t$ and $f^\text{pub}_t$; the acting agent then selects the action $u^a_t = \pidet (f^{a})$ by supplying the private observation hidden from the public agent; the public agent then uses the observed action $u^a_t$ to construct the new belief $\pubB_{t+1}$. 

\subsection{Public Belief MDP} Since $\pidet $  and  $u^a_t$ are public information, observing $u^a_t$ induces a posterior belief over the possible private state features $f_t^\text{pri}$ given by the \emph{public belief update}:
\begin{align}
 	\label{eq:belief_update}
P(f_t^{a} |  u^a_t, \pubB_t,  f_t^\text{pub}, \pidet )
&= 	\frac{P( u^a_t | f_t^{a}, \pidet )P(f_t^{a} | \pubB_t,  f_t^\text{pub}) }{  P( u^a_t | \pubB_t, f_t^\text{pub}, \pidet) } \\
&\propto  \mathbbm{1}(\pidet (f_t^{a}), u^a_t) P( f_t^{a} | \pubB_t,  f_t^\text{pub}). 
\end{align}
Using this Bayesian belief update, we can define a new Markov process, the \emph{public belief MDP} (PuB-MDP), as illustrated in Figure~\ref{fig:fig_1}b. 
The state $s_\text{BAD} = \{\pubB, f^\text{pub} \}$ of the PuB-MDP consists of the public observation and public belief; the action space is the set of deterministic partial policies that map from private observations to environment actions; and the transition function is given by $P(s'_\text{BAD} |s_\text{BAD}, \pidet )$. The next state contains the new public belief calculated using the public belief update. %
The reward function marginalises over the private state features: 
\begin{align}
	&r_\text{BAD}(s_\text{BAD}, \pidet )
	= \sum_{f^\text{pri}} \pubB(f^\text{pri}) r(s, \pidet (f^\text{pri})).
\end{align}
 Since $s'_\text{BAD}$ includes the new public belief, and that belief is computed via an update that conditions on $\pidet $, the PuB-MDP transition function conditions on all of $\pidet$, not just the selected action $u^a_t$. 
  Thus the state transition depends not just on the executed action, but on the \emph{counterfactual actions}: those specified by $\pidet $ for private observations other than $f^a_t$.

In the remainder of this section, we describe how factorised beliefs and policies can be used to efficiently learn a public policy $\pi_\text{BAD}$ for the PuB-MDP.

\subsection{Sampling Deterministic Partial Policies} For each public state,  $\pi_{\text{BAD}}$ must select a distribution $\pi_{\text{BAD}}(\pidet  |s_\text{BAD})$ over deterministic partial policies. The size of this space is exponential in the number of possible private observations $|f^a|$, but we can reduce this to a linear dependence by assuming a distribution across $\pidet$ that factorises across the different private observations, i.e., for all $\pidet$,
\begin{align}
	\pi_{\text{BAD}}(\pidet | \pubB_t, f^\text{pub}) 
&\vcentcolon= \prod_{f^a} \pi_{\text{BAD}}(\pidet(f^a) | \pubB_t, f^\text{pub}, f^a).
\end{align} 
With this restriction, we can parameterise $\pi_{\text{BAD}}$ with factors of the form $\pi^\theta_\text{BAD}(u^a | \pubB_t, f^\text{pub}, f^a)$ using a function approximator such as a deep neural network.
In order for all of the agents to perform the public belief update, the sampled $\pidet$ must be public.  We resolve this by having $\pidet$ sampled deterministically from a given $\pubB_t$ and $f_t^\text{pub}$, using a common knowledge random seed $\xi_t$.  The seeds are shared prior to the game so that all agents sample the same $\pidet$. This resembles the way humans share common ways of reasoning in card games and allows the agents to explore alternative policies jointly as a team. Further details on the mechanics of parameterising and sampling from $\pi_{\text{BAD}}$ are provided in the Supplemental Material.

\subsection{Factorised Belief Updates.} In general, representing exact beliefs is intractable in all but the smallest state spaces. For example, in card games the number of possible hands is typically exponential in the number of cards held by all players. 
To avoid this unfavourable scaling, we can instead represent an approximate factorised belief state
\begin{equation}
P( f^\text{pri}_t  | f^\text{pub}_{\leq t}) \approx 
\prod_{i }  P( f^\text{pri}_t[i] | f^\text{pub}_{\leq t}) =\vcentcolon \pubB^\text{fact}_t .
\end{equation}

\begin{figure}[]
 	\centering
		\includegraphics[width=0.95\linewidth]{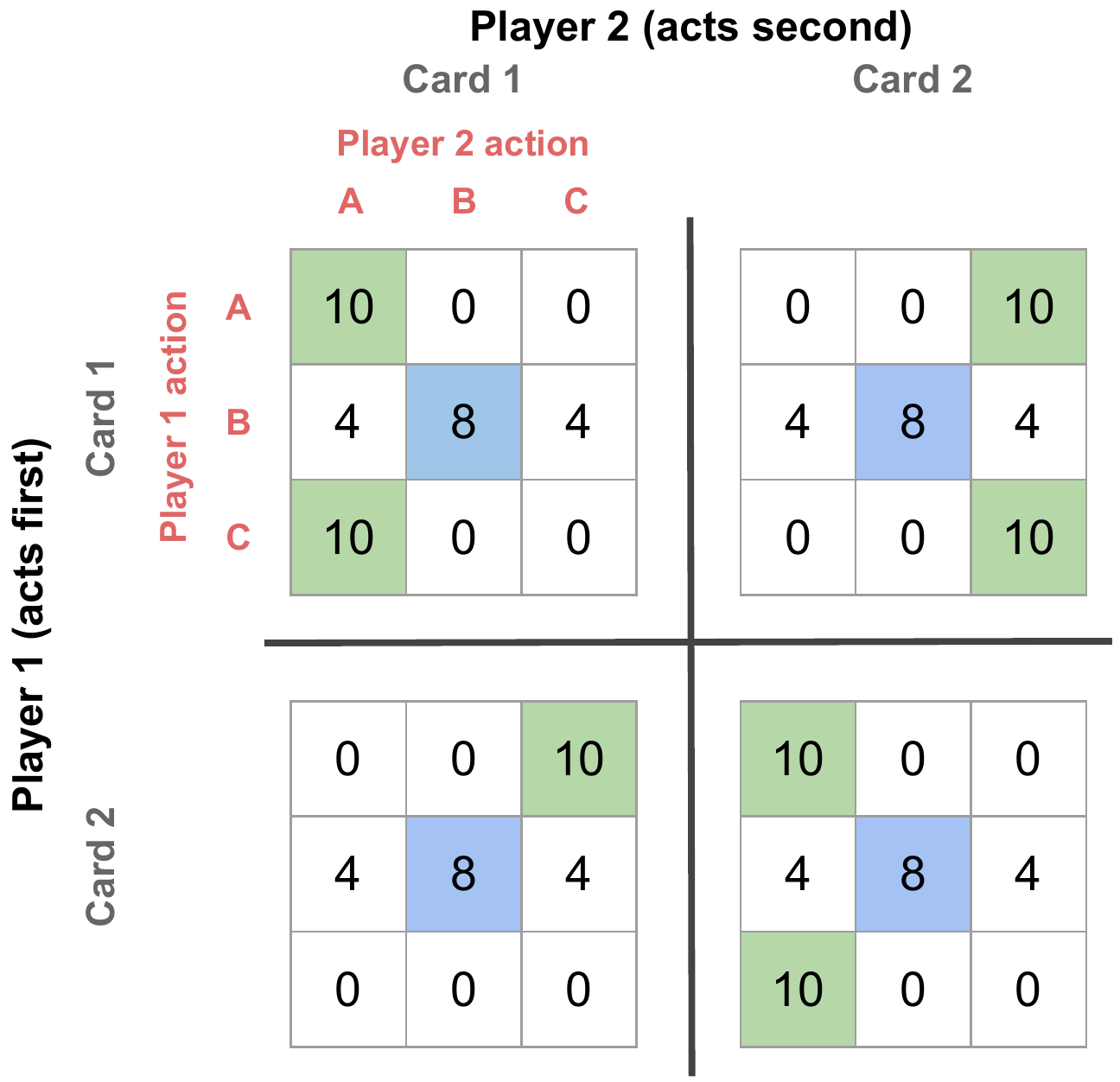}
  	\caption{Payoffs for the toy matrix-like game. The two outer dimensions correspond to the card held by each player, the two inner dimensions to the action chosen by each player. Payouts are structured such that Player 1 must encode information about their card in the action they chose in order to obtain maximum payoffs. Although presented here in matrix form for compactness, this is a two-step, turn-based game, with Player 1 always taking the first action and Player 2 taking an action after observing Player 1's action.}
 	\label{fig:tiny_hanabi}
 \end{figure}
 
From here on we drop the superscript and use $\pubB$ exclusively to refer to the factorised belief.
In a card game each factor represents per-card probability distributions, 
 assuming approximate independence across the different cards both within a hand and across players. This approximation makes it possible to represent and reason over the otherwise intractably large state spaces that commonly occur in many settings, including card games.

To carry out the public belief update with a factorised representation we maintain factorised likelihood terms ${\cal L}_t[f[i]]$ for each private feature that we update recursively:
\begin{align}
{\cal L}_t[f[i]] &\vcentcolon= P(u^a_{\leq t}|\fei, \pubB_{\leq t},  f_{\leq t}^\text{pub}, \pidet _{\leq t})  \\
&\approx {\cal L}_{t-1}[\fei] \cdot P(u^a_t | \fei, \pubB_{t},  f_{t}^{\text{pub}}, \pidet _{t}) \label{eqn:likelihood-recurse}\\
&= {\cal L}_{t-1}[\fei] \cdot 
\frac{ \mathbb{E}_{f_t \sim \pubB_t} \big[ \mathbbm{1} (\ftei, \fei) \mathbbm{1} (\pidet ( f^{a}_t), u^a_t) \big] }{ 
  \mathbb{E}_{f_t \sim \pubB_t} \big[ \mathbbm{1} (\ftei, \fei) \big]}, \label{eqn:bad-sampling}
\end{align}
where \eqref{eqn:likelihood-recurse} assumes that actions are (approximately) conditionally independent of the future given the past. As indicated, these likelihood terms are calculated by sampling, and the more samples the better.

\subsection{Self-Consistent Beliefs} 
This factorisation is only an approximation, even in  simple card games: knowledge that a player is holding a specific card clearly influences the probability that another player is holding that same card.  Furthermore, our approximation can yield beliefs that are not even self-consistent, i.e., they are not the marginalisation of any belief over joint features.  While not central to the key ideas behind BAD, we introduce a general iterative procedure that can account for feature interactions in factorised models.  Starting with a public belief $\pubB^0=\pubB_t$ we can iteratively update the belief to make it more self-consistent through re-marginalisation:
\begin{align}
\label{eq:self_consistent}
\pubB^{k+1}(\fei) &=  \sum_{\fmi}  \pubB^k(\fmi) P(\fei|\fmi,  f_{\leq t}^\text{pub}, u^a_{\leq t},  \pidet _{\leq t}) \\
								  &\propto  \mathbb{E}_{\fmi \sim \pubB^k}  \big[  {\cal L}_t(\fei) P(\fei|\fmi, f_t^\text{pub}) \big], \label{eq:B_kp1}
\end{align}
where $\fmi$ denotes all features excluding $\fei$. In the last step we used the factorised likelihoods from above to convert to an expectation, so that we can use samples to approximate the intractable sum across features. The notion of refining the distribution over one feature while keeping the distribution across all other features fixed is similar to Expectation Propagation for factor graphs~\cite{Minka2001}. However, the card counts constitute a global factor, making the factor graph formulation less useful.
While this iterative update can in principle be carried out until convergence, in practice we terminate after a fixed number of iterations.

\section{Experiments and Results}
\label{sec:experiments}

\subsection{Matrix Game}
We first present proof-of-principle results for a two-player, two-step partially observable matrix-like game (Figure~\ref{fig:tiny_hanabi}). The state consists of 2 random bits (the cards for Player 1 and 2) and the action space consists of 3 discrete actions. 
Each player observes its own card, with Player 2 also observing Player 1's action before acting, which in principle allows Player 1 to encode information about its card with its action.
The reward is specified by a payoff tensor, $r = \mathrm{Payoff}[\text{card}^1][\text{card}^2][u^1][u^2]$, where $\text{card}^a$ and $u^a$ are the card and action of the two players, respectively. The payout tensor is structured such that the optimal reward can only be achieved if the two players establish a convention, in particular if Player 1 chooses informative actions that can be decoded by Player 2.

\begin{figure}[t!]
 	\centering
	
	\begin{subfigure}[b]{0.95\linewidth}
		\includegraphics[width=\linewidth]{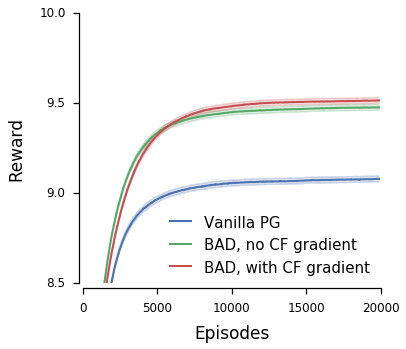}
 	\end{subfigure}
	 	\caption{BAD, both with and without counterfactual (CF) gradients, outperforms vanilla policy gradient on the matrix-like game. Each line is the mean over 1000 games, and the shade indicates the standard error of the mean (s.e.m.).}
 	\label{fig:matrix_game_result}
 \end{figure}

As shown in Figure~\ref{fig:matrix_game_result}, BAD clearly outperforms the baseline policy-gradient method on the toy matrix game. In this small, exact setting, it is also possible to estimate counterfactual (CF) policy gradients that reinforce not only the action taken, but also these counterfactual actions. This can be achieved by replacing $\log \pi^a( u^a_t | \tau^a) $ with $\log P(\pidet  |  \pubB_t,  f^\text{pub})$ in the estimation of the policy gradient. However, the additional improvement in performance from using CF gradients is minor compared to the initial performance gain from  using a counterfactual belief state. 

Code for the matrix game with a proof-of-principle implementation of BAD is available at \url{https://bit.ly/2P3YOyd}.

\subsection{Hanabi}
Here we briefly describe the rules of 2-player Hanabi. 

There are 5 cards in a hand. For each of the 5 colours there are three $\textbf{1}$s, one $\textbf{5}$, and two each of all other ranks, i.e., $10$ cards per colour for a total of $50=5 \times 10$ cards in the deck.

While this is a modestly large number of cards, even for 2 players it leads to $6.2 \times 10^{13}$ possible joint hands at the beginning of the game. 

\subsection{Observations and Actions}
Each player observes the hands of all other players, but not their own. The action space consists of $ 2 \times 5$ options for discarding and playing cards, and $5 + 5$ options per teammate for hinting colours and ranks. Hints reveal all cards of a specific rank or colour to one of the teammates, e.g., `Player 2's card 3 and 5 are red'. Hinting for colours and ranks not present in the hand of the teammate (so-called `empty hints') is not allowed.

Each hint costs one hint token. The game starts with $8$ hint tokens, which can be recovered by discarding cards. After a player has played or discarded a card, she draws a new card from the deck. When a player picks up the last card, everyone (including that player) takes one more action before the game terminates.
Legal gameplay consists of building $5$ fireworks, which are piles of ascending numbers, starting at $\textbf{1}$, for each colour. When the $\textbf{5}$ has been added to a pile the firework is complete and the team obtains another hint token (unless they already have 8).
A life token is lost each time a player plays an illegal card; after three mistakes the game terminates. 
Players receive 1 point after playing any playable card, with a perfect score being $25 = 5 \times 5$.

The number of hint and life tokens at any time are observed by all players, as are the played and discarded cards, the last action of the acting player and any hints provided.

\subsection{Beliefs in Hanabi}
\label{sec:beliefs_in_hanabi}
The basic belief calculation in Hanabi is straightforward: $f_t^\text{pub}$ consists of a vector of `candidates' $C$ containing counts for all remaining cards, and a `hint mask' $\text{HM}$, an $AN_h\times (N_\text{color}N_\text{rank}+1)$ binary matrix that is $1$ if in a given `slot' the player could be holding a specific card according to the hints so far, and $0$ otherwise; the additional 1 accounts for the possibility that the card may not exist in the final round of play. Slots correspond to the features of the private state space $\fei$, for example the 3rd card of the second player. Hints contain both positive and negative information: for example, the statement `the 2nd and 4th cards are red' also implies that all other cards are not red.

The basic belief $B^0$ can be calculated as
\begin{equation}
B^0(\fei)= P(\fei | f^\text{pub} ) \propto C(f) \times \text{HM}(\fei).
\end{equation}
 We call this the `V0 belief', in which the belief for each card depends only on publicly available information for that card. In our experiments, we focus on baseline agents that receive this basic belief, rather than the raw hints, as public observation inputs; while the problem of simply remembering all hints and their most immediate implication for card counts is potentially challenging for humans in recreational play, we are here more interested in the problem of forming effective conventions for high-level play.
  
As noted above, this basic belief misses an important interaction between the hints for different slots.  
We can calculate an approximate version of the self-consistent beliefs that avoids the potentially expensive and noisy sampling step in Equation~\ref{eq:B_kp1} (note that this sampling is distinct from the sampling required to compute the marginal likelihood in Equation~\ref{eqn:bad-sampling}). As derived in the Supplemental Material,
\begin{equation}
B^{k+1}(\fei) \propto \Bigg( C(f)  - \sum_{j \neq i} B^{k}(\fej) \Bigg)   \times \text{HM}(\fei).
\end{equation}
We call the resulting belief at convergence (or after a maximum number of iterations) the `V1 belief`. It does not condition on the Bayesian probabilities but considers interactions between hints for different cards. In essence, at each iteration the belief for a given slot is updated by reducing the candidate count by the number of cards believed to be held across all other slots.
 
By running the same algorithm but including ${\cal L}$, we obtain the Bayesian beliefs $\text{BB}$ that lie at the core of BAD:
\begin{align}
	\text{BB}^0 (\fei) & \propto C(f)  \times \text{HM}(\fei) \times {\cal L}(\fei),  \\
\text{BB}^{k+1}(\fei) &\propto \Bigg( C(f)   - \sum_{j \neq i} B^{k}(\fej) \Bigg) \notag \\ &\qquad \times  \text{HM}(\fei) \times {\cal L}(\fei).
\end{align}
In practice, to ensure stability, the final `V2 belief' that we use is an interpolation between the Bayesian belief and the V1 belief: $\text{V2} = (1 - \alpha) \text{BB} +  \alpha \text{V1}$ with $\alpha=0.01$ (we found $\alpha=0.1$ to also work). 
For the Bayesian update we sampled $S=3,000$ hands during training and $S=20,000$ hands for the final test games. 

  \begin{figure*}[t!]
 	\centering
	
 	\begin{subfigure}[b]{0.32\linewidth}
		\includegraphics[width=\linewidth]{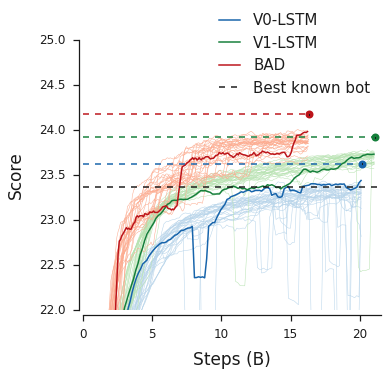}
		\caption{}
 	\end{subfigure}
	\begin{subfigure}[b]{0.32\linewidth}
  		\includegraphics[width=\linewidth]{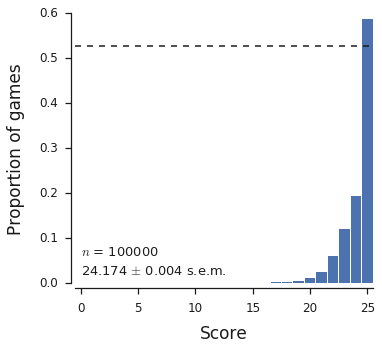}
		\caption{}
 	\end{subfigure}
	\begin{subfigure}[b]{0.32\linewidth}
  		\includegraphics[width=\linewidth]{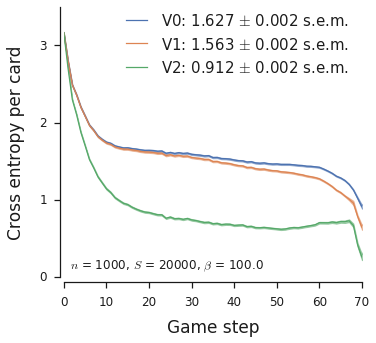}
		\caption{}
 	\end{subfigure}
	 	\caption{a) Hanabi training curves for BAD and the V0 and V1 baseline methods using LSTMs rather than the Bayesian belief. Thick lines indicate the final evaluated agent for each agent type, with the dots showing the final test score. Error bars (standard error of the mean, s.e.m.) are smaller than the dots. Upward kinks in the curves are generally due to agents `evolving' in PBT by copying its weights and hyperparameters (plus perturbations) from a superior agent.
 	b) Distribution of game scores for BAD on Hanabi under testing conditions. BAD achieves a perfect score in almost $60\%$ of the games. The dashed line shows the proportion of perfect games reported for FireFlower, the best known hard-coded bot for two-player Hanabi. c) Per-card cross entropy with the true hand for different belief mechanisms during BAD play. V0 is the basic belief based on hints and card counts, V1 is the self-consistent belief, and V2 is the BAD belief which also includes the Bayesian update. The BAD agent conveys around $40\%$ of the information via conventions, rather than grounded information.}
 	\label{fig:full_game}
 \end{figure*}

\subsection{Architecture Details for Baselines and Method}
Advantage actor-critic agents were trained using the Importance-Weighted Actor-Learner Architecture \cite{Espeholt2018}, in particular the multi-agent implementation described in \OnlineCite{Jaderberg2018}. In this framework, `actors' continually generate trajectories of experience (sequences of states, actions, and rewards) by having agents (self-)playing the game, which are then used by `learners' to perform batched gradient updates (batch size was 32 for all agents). Because the policy used to generate the trajectory can be several gradient updates behind the policy at the time of the gradient update, V-trace was applied to correct for the off-policy trajectories. The length of the trajectories, or rollouts, was 65, the maximum length of a winning game.

Further details for the hyperparameters, architecture, and training are given in the Supplemental Material.

\subsection{Results on Hanabi}

The BAD agent achieves a new state-of-the-art mean performance of \finalperformance{} points on two-player Hanabi. In Figure~\ref{fig:full_game}a we show training curves and test performance for BAD and two LSTM-based baseline methods, as well as the performance of FireFlower (\url{https://github.com/lightvector/fireflower}), the best known hand-coded bot for two-player Hanabi. For the LSTM agents, test performance was obtained by using the greedy version of the trained policy, resulting in slightly higher scores than during training. To select the agent, we first performed a sweep over all agents for 10,000 games, then carried out a final test run of 100,000 games on the best agent from the sweep to obtain an unbiased score. For the BAD agent we also increased the number of sampled hands. 
The results for other learning methods from the literature perform below the range of the $y$-axis (far below 20 points) and are omitted for readability. We note that, under a strict interpretation of the rules of Hanabi, games in which all three error tokens are exhausted should be awarded a score of 0. Under these rules the same BAD agent achieves $23.917 \pm 0.009$ s.e.m, the best known score, even though it was not trained under these conditions. SmartBot and FireFlower achieve average scores of 22.99 and 22.56 respectively. 
 \begin{table}[t!]
	\begin{center}
		\resizebox{\linewidth}{!}{
			\begin{tabular}{c ccc}
				\toprule
				
				\multirow{2}{*}{Agent} 
				& \multirow{2}{*}{Learning steps} 
				& \multirow{2}{*}{Mean $\pm$ s.e.m.}
				& \multirow{2}{*}{Prop. perfect}
				\\
				&  &  &  \\

				\midrule
				SmartBot & - & 23.09 & 29.52\% \\
				FireFlower & - & 23.37 $\pm$ 0.0002 & 52.6\% \\
				V0-LSTM & 20.2B  & 23.622 $\pm$ 0.005 & 36.5\% \\
				V1-LSTM & 21.1B  & 23.919 $\pm$ 0.004 & 47.5\% \\
				BAD & \finalsteps B & \textbf{\finalperformance $\pm$ 0.004} & \textbf{\finalperfectgames\%} \\
				\bottomrule

			\end{tabular}
		}
	\end{center}
	\caption{Test scores on 100K games. The LSTM agents were tested with a greedy version of the trained policy, while the final BAD agent was evaluated with V1 mix-in $\alpha=0.01$, 20K sampled hands, and inverse softmax temperature 100.0. The FireFlower bot was evaluated over 25K games.}
	\label{tbl:scores}
	
\end{table}

While not all of the game play BAD learns is easy to follow, some conventions can be understood simply from inspecting the game. Printouts of 100 random games can be found at \url{https://bit.ly/2zeEShh}. One convention stands out:  Hinting for `red' or `yellow' indicates that the newest card of the other player is playable. We found that in over $80\%$ of cases when an agent hints `red' or `yellow',  the next action of the other agent is to play the newest card.
This convention is very powerful: Typically agents know the least about the newest card, so by hinting `red' or `yellow', agents can use a single hint to tell the other agent that the card is playable. Indeed, the use of two colours to indicate `play newest card' was present all of the highest-performing agents we studied. Hinting `white' and `blue' are followed by a discard of the newest card in over $25\%$ of cases. We also found that the agent sometimes attempts to play cards which are not playable in order to convey information to their team mate. 
In general, unlike human players, agents play and discard predominantly from the last card. In 
the supplementary material we also include a written analysis of our bot by the creator of FireFlower.

Figure~\ref{fig:full_game}c shows the quality of the different beliefs. While the iterated belief update leads to a reduction in cross entropy compared to the basic belief, a much greater reduction in cross entropy is obtained using counterfactual beliefs. This clearly demonstrates the importance of learning conventions for successful gameplay in Hanabi: Roughly $40\%$ of the information is obtained through conventions rather than through the grounded information and card counting.

\section{Related Work}
\label{sec:related_work}

\subsection{Learning to Communicate} Many works have addressed problem settings where agents must learn to communicate in order to cooperatively solve a toy problem. 
These tasks typically involve a cheap-talk communication channel that can be modeled as a continuous variable during training, which allows differentiation through the channel as first proposed by \OnlineCite{foerster2016learning} and \OnlineCite{Sukhbaatar2016}. In this work we focused on the case where agents must learn to communicate via grounded hinting actions and observable environment actions rather than a cheap-talk channel. This is closest to the ``hat game'' of \OnlineCite{foerster2016learning1}, who proposed a simple extension to recurrent deep $Q$-networks rather than explicitly modeling action-conditioned Bayesian beliefs.  %
An idea similar to the Pub-MDP was introduced in the context of decentralised stochastic control by \OnlineCite{Nayyar2013}, who also formulated a coordinator that uses ``common information'' to map local controller information to actions. However, they did not provide a concrete solution method that can scale to a high-dimensional problem like Hanabi. 
\subsection{Hanabi} A number of papers have been published on Hanabi. \OnlineCite{baffier2016hanabi} showed that optimal gameplay in Hanabi is NP-hard even when players can observe their own cards. Encoding schemes similar to the hat game essentially solve the five-player case~\Cite{Cox2015}, but only achieve 17.8 points in the two-player setting \Cite{bouzy2017playing}. \OnlineCite{walton2017evaluating} developed a variety of Monte Carlo tree search and rule-based methods for Hanabi, but the reported scores were roughly $50\%$ lower than BAD. \OnlineCite{osawa2015solving} defined a number of heuristics for the two-player case that reason over possible hands given the other player's action. While this is similar in spirit to our approach, the work was limited to hand-coded heuristics, and the reported scores were around 8 points lower than BAD. \OnlineCite{Eger2017} investigated humans playing with hand-coded agents, but no pairing resulted in scores higher than 15 points on average.

The best result for two-player Hanabi we could find was for the FireFlower described at \url{github.com/lightvector/fireflower}, which has been reported to achieve an average of $23.37$ points (52.6\% perfect games). While FireFlower uses the same game rules as those used in our work, it is entirely hand-coded and involves no learning.

\subsection{Belief State Methods} The continual re-solving (nested solving) algorithm used by DeepStack~\Cite{moravvcik2017deepstack}  and Libratus~\Cite{brown2018superhuman} for poker also use a belief state space. Like BAD, when making a decision in a player state, continual re-solving considers the belief state associated with the current player and generates a joint policy across all player states consistent with this belief. The policy for the actual player is then selected from this joint policy. Continual re-solving also does a Bayesian update of the beliefs after an action. There are key differences, however. Continual re-solving performa exact belief updates, which requires a joint policy space  small enough to enumerate; belief states are also augmented with opponent values; continual re-solving is a value-based method, where the training process consists of learning the values of belief states under optimal play; finally, the algorithm is designed for two-player, zero-sum games, where it can independently consider player state values while guaranteeing that an optimal choice for the joint action policy can be found.

\section{Conclusion and Future Work}
We presented the \emph{Bayesian action decoder} (BAD), a novel algorithm for multi-agent reinforcement learning in cooperative partially observable settings. BAD uses a factorised, approximate belief state that allows agents to efficiently learn informative actions, leading to the discovery of conventions. 
We showed that BAD outperforms policy gradients in a proof-of-principle matrix game, and achieves a state-of-the-art performance of \finalperformance{} points on average in the card game Hanabi. We also showed that using the Bayesian update leads to a reduction in uncertainty across the private hands in Hanabi by around $40\%$.  To the best of our knowledge, this is the first instance in which deep RL has been successfully applied to a problem setting that both requires the discovery of communication protocols and was originally designed to be challenging for humans. BAD also illustrates clearly that using an explicit belief computation achieves better performance in such settings than current state-of-the-art RL methods using implicit beliefs, such as recurrent neural networks. 

In the future, we aim to apply BAD to games with more players and further generalise BAD by learning more of its components, e.g., the V0-belief. While the belief update necessarily involves a sampling step, most of the other components can likely be learned end-to-end. 
We also plan to extend the BAD mechanism to value-based methods and further investigate the relevance of counterfactual gradients. Similar to what was suggested as next steps in~\cite{zinkevich2011lemonade}, we hope to extend the setting to a point where our bots can learn to collaborate with human players. 


\section*{Acknowledgements}

We thank Marc Lantot, Shibl Mourad, Angeliki Lazaridou, Jelena Luketina, Anuj Mahajan, Gregory Farquhar, Kelsey Allen, Thore Graepel, Nando de Freitas, and Nolan Bard for valuable discussions.

This project has received funding from the European Research Council (ERC) under the European Union’s Horizon 2020 research and innovation programme (grant agreement \#637713).


\bibliography{include/hanabi}
\bibliographystyle{include/icml2018/icml2018}

\clearpage
\newpage

\appendix

\section{Parameterising and Sampling from the Distribution over Partial Policies}
\label{supp:samplePartial}

BAD requires us to parameterise a probability distribution over partial policies using a deep neural network:
\begin{align}
	P(\pidet | s_\text{BAD}) = \pi^\theta_\text{BAD}(\pidet | s_\text{BAD}).
\end{align}

The first insight is that we can trivially use a neural network to map from public states, $ s_\text{BAD}$, into probabilistic partial policies.
To do so, we simply start with a feedforward policy that takes as input both  $ s_\text{BAD}$ and $f^a$ and produces a distributions over actions:
\begin{align}
 \pi^\theta( s_\text{BAD}, f^a 	) \rightarrow P(u |  s_\text{BAD}, f^a).
\end{align}

Next, we note that if we fix a given $ s_\text{BAD}$, we now have a probabilistic partial policy which maps each private observation $f^a$ into a probability distribution over actions. This partial policy is produced deterministically as a function of  $s_\text{BAD}$ via the parameters $\theta$:

\begin{align}
\pi(u|f^a): \{f^a\} &\rightarrow \{P(\mathcal{U})\} \;  | s_\text{BAD}, \\
 \pi(u|f^a)  &= \pi^\theta( u|s_\text{BAD}, f^a 	).
 \end{align}
 
 Now, this is close, but not quite what we want. Above we have a deterministic map from $s_\text{BAD}$ into probabilistic partial policies, $\pi(u|f^a)$. Instead, we require a differentiable distribution over deterministic partial policies. 
 
 Perhaps surprisingly, this can be accomplished by conditioning the sampling from $\pi(u|f^a)$ on a common knowledge random seed, $\xi$:
 \begin{align}
 \pidet: \{f^a \} &\rightarrow \mathcal{U} \; |  s_\text{BAD}, \\
  \{f^a \} &\rightarrow u \sim \pi(u|f^a) \;  | \xi, \\
   \{f^a \} &\rightarrow u \sim \pi^\theta(u| s_\text{BAD}, f^a 	)  \;  | \xi. \\
    \end{align}

Thus, when we sample $\xi$ we are effectively sampling an entire deterministic partial policy.


\section{Hyperparameters and Training Details}
\label{sec:hyperparams}

For the toy matrix game, we used a batch size of 32 and the Adam optimiser with all default TensorFlow settings; we did not tune hyperparameters for any runs.

 In the V0-LSTM and V1-LSTM BAD agents, all observations were first processed by an MLP with a single 256-unit hidden layer and ReLU activations, then fed into a 2-layer LSTM with 256 units in each layer. The policy $\pi$ was a softmax readout of the LSTM output. The baseline network was an MLP with a single 256-unit hidden layer and ReLU activations, which then projected linearly to a single value. Since the baseline network is only used to compute gradient updates, we followed \citeauthor{Foerster2017} (\citeyear{Foerster2017}) in feeding each agent's own hand (i.e., the other agent's private observation) into the baseline by concatenating it with the LSTM output; thus we make the common assumption of centralised training and decentralised execution. We note that the V0 and V1-LSTM agents differed \emph{only} in their public belief inputs.

  The Hanabi BAD agent consisted of an MLP with two 384-unit hidden layers and ReLU activations that processed all observations, followed by a linear softmax policy readout. To compute the baseline, we used the same MLP as the policy but included the agent's own hand in the input (this input was present but zeroed out for the computation of the policy).

For all agents, illegal actions (such as hint for a red card when there are no red cards) were masked out by setting the corresponding policy logits to a large negative value before sampling an action. In particular, for the non-acting agent at each turn the only allowed action was the `no-action'.
For Hanabi, we used the RMSProp optimiser with $\epsilon=10^{-10}$, momentum 0, and decay 0.99. The RL discounting factor $\gamma$ was set to 0.999. The baseline loss was multiplied by 0.25 and added to the policy-gradient loss. We used population-based training (PBT) \cite{Jaderberg2017,Jaderberg2018} to `evolve' the learning rate and entropy regularisation parameter during the course of training, with each training run consisting of a population of 30 agents.
 For the LSTM agents, learning rates were sampled log-uniformly from the interval $[1, 4)\times10^{-4}$ while the entropy regularisation parameter was sampled log-uniformly from the interval $[1, 5)\times10^{-2}$. For the BAD agents, learning rates were sampled log-uniformly from the interval $[9\times10^{-5},\ 3\times10^{-4})$ while the entropy regularisation parameter was sampled log-uniformly from the interval $[3, 7)\times10^{-2}$. Agents evolved within the PBT framework by copying weights and hyperparameters (plus perturbations) according to each agent's rating, which was an exponentially moving average of the episode rewards with factor 0.01. An agent was considered for copying roughly every 200M steps if a randomly chosen copy-to agent had a rating at least 0.5 points higher. To allow the best hyperparameters to manifest sufficiently, PBT was turned off for the first 1B steps of training.
 
The BAD agent was trained with 100 self-consistent iterations, a V1 mix-in of $\alpha=0.01$, BAD discount factor $\gamma_\text{BAD}=1$, inverse temperature 1.0, and 3000 sampled hands. Since sampling from card-factorised beliefs can result in hands that are not compatible with the deck, we sampled 5 times the number of hands and accepted the first 3000 legal hands, zeroing out any hands that were illegal.

\section{Self-Consistent Belief Approximation for Hanabi}
\label{sec:iterativeproof}
We will use the same notation as in the main text: $f_t^\text{pub}$ consists of a vector of `candidates' $C$ containing counts for all remaining cards, and a `hint mask' $\text{HM}$, an $AN_h\times N_\text{color}N_\text{rank}$ binary matrix that is $1$ if in a given `slot' the player could be holding a specific card according to the hints given so far, and $0$ otherwise''.
Furthermore, ${\cal L}(\fei) $, is the marginal likelyhood.

Then the basic per-card belief is simply:
 \begin{align}
B^0(\fei)&\propto C(f) \times \text{HM}(\fei) \times {\cal L}(\fei),  \\
B^0(\fei) &= \frac{C(f) \times \text{HM}(\fei) \times {\cal L}(\fei)}{ \sum_{g}  C(g) \times \text{HM}(\gei) \times {\cal L}(\gei)}   \\
&= \beta_i \big( C(f) \times \text{HM}(\fei) \times {\cal L}(\fei)  \big). 
 \end{align}
 
 In the last two lines we are normalising the probability, since the probability of the $i$-th feature being one of the possible values must sum to $1$. For convenience we also introduced the notation $\beta_i$ for the normalisation factor.
 
 Next we apply the same logic to the iterative belief update. The key insight here is to note that conditioning on the features $\fmi$, i.e., the other cards in the slots, corresponds to reducing the card counts in the candidates. Below we use $M(\fei) = \text{HM}(\fei) \times {\cal L}(\fei)  $ for notational convenience:
 \begin{align}
&\pubB^{k+1}(\fei) \notag \\
&=\sum_{\fmi }  \pubB^k(\fmi) P(\fei|\fmi,  f_{\leq t}^\text{pub}, u^a_{\leq t}, \pidet _{\leq t})  \\
 &=  \sum_{\gmi }  \pubB^k(\gmi) \beta_i \bigg( C(f)-\sum_{j \neq i} \mathbbm{1}(\gej = f) \bigg )M(\fei).
\end{align}

In the last line we relabelled the dummy index $\fmi$ to $\gmi$ for clarity and used the result from above.
Next we substitute the factorised belief assumption across the features, $ \pubB^k(\gmi)  = \prod_{j \neq i} \pubB^k(\gej) $ :
 \begin{align}
 &\pubB^{k+1}(\fei) \notag \\
 &=  \sum_{\gmi }  \pubB^k(\gmi) \beta_i \bigg( C(f)-\sum_{j \neq i} \mathbbm{1}(\gej = f)\bigg)M(\fei)  \\
  &=  \sum_{\gmi }   \prod_{j \neq i} \pubB^k(\gej)  \beta_i \bigg( C(f)-\sum_{j \neq i} \mathbbm{1}(\gej = f)   \bigg)M(\fei)  \\
 &\simeq \beta_i   \sum_{\gmi }   \prod_{j \neq i} \pubB^k(\gej)  \bigg( C(f)-\sum_{j \neq i} \mathbbm{1}(\gej = f) \bigg)M(\fei).
\end{align}

In the last line we have ommited the dependency of $\beta_i$ on the sampled hands $\fmi$. It corresponds to calculating the average across sampled hands first and then normalising (which is approximate but tractable) rather than normalising and then averaging (which is exact but intractable). We can now use product-sum rules to simplify the expression. 
\begin{align}
&\pubB^{k+1}(\fei) \notag \\
&\simeq \beta_i    \bigg( C(f)-  \sum_{\gmi }   \prod_{j \neq i} \pubB^k(\gej) \sum_{j \neq i} \mathbbm{1}(\gej = f) \bigg)M(\fei)  \\	
&= \beta_i    \bigg( C(f)-  \sum_{j \neq i }  \sum_{g}  \pubB^k(\gej) \mathbbm{1}(\gej = f) \bigg)M(\fei)  \\	
&= \beta_i    \bigg(C(f)-  \sum_{j \neq i }  \pubB^k(\fej)   \bigg)M(\fei)  \\	
&\propto    \bigg(C(f)-  \sum_{j \neq i }  \pubB^k(\fej)   \bigg)M(\fei).
\end{align}

This concludes the proof.

\section{Anecdotal Analysis}
\label{app:anedotal}
Below we present commentary from David Wu (\url{https://github.com/lightvector/}), the creator of the FireFlower bot, on our BAD agent.  
While this is anecdotal evidence, we believe it provides some interesting insights into the gameplay that our BAD agent discovers. The comments are taking verbatim from an email exchange with David:

\subsection{Communicating Playables}
\begin{itemize}
	\item As you observed before, the bot uses R and Y often to hint newest-card-playability.
\item In addition to the R and Y hints, it also often uses direct hints to the newest card to indicate playability, in the way that natural human conventions do, and I think these include both color and number hints.
\item When the R and Y hints or direct hints to the newest card hit multiple cards, the bot often was indicating multiple plays. In the small sample size of cases we looked over, it tended to be the case that the R/Y hints were more often ``play in the order from newest to oldest'' while the direct hints were more often in the order of ``play from oldest to newest''. I think this was not 100\% consistent though, but in all cases when looking at the direct beliefs, it was clear that in each case there was a strong ordering convention was in force for that hint, it's just that we didn't see enough cases to be able to determine the precise rules for which one when. Generally though, it makes a lot of sense to vary the ordering convention in different parts of the hint space to add flexibility in hinting.
\item The bot uses certain other kinds of direct hints to older cards to suggest that those cards are one step away from playable, or something of that nature. Sometimes the belief state shows that this is not absolutely certain, but over time as other things happen the probability mass sometimes gradually updates and concentrates on the card on the truth, such that once the preceding card is played, the bot may then play the formerly-one-step-away card without any further suggestion.
\item For these ``delayed'' one-step-removed hints to older cards, there is also a similar variation in ordering conventions in the case those hints hit more than one card, sometimes they're in ``age-order'' and sometimes they're in ``reverse-age-order''. 
\item Commonly the R and Y hints also indicate other plays or delayed plays besides the play of the newest card. The bot chooses the manner of hinting the first card  as playable (R vs Y vs direct hint) to try to communicate other useful information at the same time, if possible.
\item I think occasionally the bot seems to ``single out'' a card by directly hinting all other cards in the hand *besides* that card over successive turns, and sometimes this implies that the singled-out unhinted card is playable. I'm not sure on this one though, I'd need to see more cases.
\item I think there seems to be some interesting other conventions that seem to function to give information to allow play of older red and yellow cards, which are necessary since direct hints of R and Y mean to play the newest card rather than the card hinted.
\end{itemize}

\subsection{Communicating Protection}
\begin{itemize}
\item As you observed before, the bot discards its newest card by default.  
\item G hints that do not directly hit the first card appear to mean that the newest card is dangerous and should not be discarded. Possibly it is more specific, and actually just means that it's a 5, the examples I recall all involved 5s. The bot also can just directly hint the newest card in various ways.
\item The bot is very aggressive about protecting the newest card if the newest card is a 5 or otherwise dangerous (the last copy in the deck), whether by giving a G hint, or a direct hint, or otherwise. This is so consistent that pretty much any action other than an immediate protection causes the other bot to infer that the newest card is NOT a 5 or the last copy of a card whose first copy has been lost.
\item However, the bot does *not* do this any longer if there is a common-knowledge-extremely-safe discard in that player's hand (e.g. a redundant copy of a card already played). In that case, it is understood that the bot will prefer to discard that instead. Then, protection of the newest card is not necessarily urgent any more, and neither will a player necessarily infer that the newest card is safe from a failure by their partner to protect it immediately.
\item There seems to be some interesting dictionary of hints that we haven't worked out yet about ways to signal to discard cards besides the newest, which prevents junk from accumulating in the hand as non-playable but useful-to-hold-on-to cards enter the hand.
\end{itemize}

Miscellaneous Communication
\begin{itemize}
\item Often the hints, and sometimes its other actions just come ``attached'' with miscellaneous information. The most extreme example is I observed one game where as a result of the bots discarding, it was immediately implied that a particular card in the other player's hand was almost certainly red. This information was not immediately useful (the red card was not yet playable, nor was it likely to have been discarded soon), it was simply just extra information attached to the action of discarding in that particular case. Presumably the bot was by convention constrained to almost certainly do some other action in that situation had that partner's card counterfactually not been red.
\item This kind of extra not-immediately-useful ``attached'' information is perhaps the most non-human part of the bot's convention set. But actually it doesn't happen as often as one might expect from a ``nonhuman'' agent. For the most part I didn't see this all that much for plays and discards (that one extreme example notwithstanding). This makes sense, as having too many such conventions would overly constrain the ability of the players to act, as discard/play are both critical actions you need to take very frequently regardless of the other player's hand. 
\item Even for hint actions, most hints were very sensible and humanly explainable, or clearly appeared that they would be humanly explainable had we had a larger sample size so that we could be surer about the generality of its meaning and exactly how the bot had packed different meanings into the hint space. There were only a few hint actions that I found particularly ``weird'' in what inference was made.
\item A priori, there's no particular reason why a bot's conventions couldn't, for example, completely change depending on the turn number modulo 3, and be extremely hard for humans to comprehend. But for the most part, the conventions of this bot weren't like that - they were pretty understandable, or at least seemed consistent and sensible even if we didn't have all the exact meanings mapped out.
\end{itemize}

Overall Quality of Play and Game Flow
\begin{itemize}
\item The bot is *very* strong in the early game, and there its convention set is overall far more efficient than ``natural'' human convention sets (although not-necessarily human convention sets that were constructed to be more artificial and encoding-like). It's really quite beautiful.

\item The bot is superhumanish at tracking inferred information over time, e.g. on the one hand inferring that a card is not scary in first position, then as it drifts back later in the hand, inferring this or that other property incidentally, and inferring based on the "aging" of the card that it is probably not this or that, and so on, until only a couple possibilities remain. It's not uncommon that in the midgame, both players know almost all the relevant things about their hands.

\item The bot might be tactically weak in occasional situations on or near 0 hints, where the ensuing sequence of actions is heavily constrained. It seems to have a very strong preference to discard and get away from 0 hints, even when as far as we can tell based on its convention set it should be possible to just stay at 0 hints and play out some cards, and where discarding at that moment is suboptimal. For example if the ensuing sequence of plays would result in a few 5s being played thereby recovering some hints for free, and the partner's immediate discard is also completely safe in the event that the partner wants to discard, whereas one's own discard unnecessarily loses a copy of a card that could be useful in the future. (If I read the paper right, there is no explicit lookahead in this bot?).

\item The bot makes a few seemingly-clear mistakes in the endgame (as far as we can tell), although only slight ones. For example, one of the games we looked at:
\begin{itemize}
   \item The players were in a close-to-winning state - they both knew all the playable cards in their hands or had inferred them with high confidence, and all they needed to do was play those cards and wait to draw the few remaining cards to play.
  \item They had plenty enough hints and headroom to theoretically execute essentially-perfect play thereafter (i.e. getting plays out in a timely manner, collectively never discarding any card that could be useful thereafter, optimizing who draws the next card for parity, etc), and by my understanding of their conventions, nothing stopping them from doing so.
  \item But instead of playing, the bot wasted a turn giving their partner a hint. When you inspected the V2 belief state, it gave no useful information - the dominant effect of the hint was actually to concentrate probability mass *away* from the truth giving the partner a misleading belief about a card, and had almost no other effects. 
   \item Their partner then proceeded to also not play and instead discarded their newest card, which unnecessarily lost one of the copies of a useful 4. There was a copy of the 4 left in the deck, but such a discard is still bad. If the remaining copy of that card is the very bottom card of the deck, it guarantees that you cannot get 25 points, so every unnecessary discard of the first copy of any card loses you EV due to the chance for the other copy to be the last card.
\end{itemize}
\end{itemize}

Speculating a little here - perhaps something about the bot's policy or convention set hasn't converged as sharply in the endgame? It's certainly the case that the gradient there is much smaller - even a clear mistake near the end tends to cost you only a little in EV if you're measuring by score, whereas near the start of the game it can cost you a lot. And the expected penalty for discarding the first copy of a useful 4 when otherwise well ahead is slight, since it then usually only harms you when that 4 is precisely the last card in the deck which only happens 1/N times, so one might imagine the average gradient there for good behavior to be very small.


\end{document}